\documentclass[12pt,a4paper,dvips]{article}
\usepackage{a4p}
\usepackage{cite,mcite}
\usepackage{graphicx}
\usepackage{physics }
\usepackage{rotating}
\usepackage{l3_title,ifthen}
\journalname{Phys. Lett. B}
\date{June 29, 2000}
\preprint{2000-084}
\DeclareGraphicsExtensions{.eps,.ps,.eps.gz,.ps.gz}
\newlength{\capindent}
\newlength{\capwidth}
\newlength{\figwidth}
\newcommand{\icaption}[2][!*!,!]{\hspace*{\capindent}%
  \begin{minipage}{\capwidth}
    \ifthenelse{\equal{#1}{!*!,!}}%
      {\caption{#2}}%
      {\caption[#1]{#2}}
  \end{minipage}}
\newcommand{\mocl}{\multicolumn{1}{c|}}
\newcommand{\mtcl}{\multicolumn{2}{c|}}

\newcommand{\mfcl}{\multicolumn{4}{c|}}

\newcommand{\ff}{\ensuremath{f\kern 0.15em\overline{\kern -0.25em f}}}

\newcommand{\EE} {\ensuremath{\mathrm{e^+e^-}}}

\newcommand{\EEFF}{\ensuremath{\rm \ee\rightarrow \mathrm{f \bar{f}}}}

\newcommand{\bitem}{\begin{itemize}}
\newcommand{\eitem}{\end{itemize}}

\newcommand{\SM}{Standard Model}

\newcommand{\MZbar}{\ensuremath{\overline{m}_{\mathrm{Z}}}}
\newcommand{\GZbar}{\ensuremath{\overline{\Gamma}_{\mathrm{Z}}}}
\newcommand{\rf}{\mathrm{r}_f}
\newcommand{\jf}{\mathrm{j}_f}
\newcommand{\gf}{\mathrm{g}_f}

\newcommand{\rtote}{\mathrm{r^{tot}_{e}}}
\newcommand{\rtotm}{\mathrm{r^{tot}_{\mu}}}
\newcommand{\rtott}{\mathrm{r^{tot}_{\tau}}}
\newcommand{\rtotl}{\mathrm{r^{tot}_{\ell}}}
\newcommand{\rtoth}{\mathrm{r^{tot}_{had}}}

\newcommand{\jtote}{\mathrm{j^{tot}_{e}}}
\newcommand{\jtotm}{\mathrm{j^{tot}_{\mu}}}
\newcommand{\jtott}{\mathrm{j^{tot}_{\tau}}}
\newcommand{\jtotl}{\mathrm{j^{tot}_{\ell}}}
\newcommand{\jtoth}{\mathrm{j^{tot}_{had}}}

\newcommand{\gtoth}{\mathrm{g^{tot}_{had}}}

\newcommand{\rfbe}{\mathrm{r^{fb}_{e}}}
\newcommand{\rfbm}{\mathrm{r^{fb}_{\mu}}}
\newcommand{\rfbt}{\mathrm{r^{fb}_{\tau}}}
\newcommand{\rfbl}{\mathrm{r^{fb}_{\ell}}}

\newcommand{\jfbe}{\mathrm{j^{fb}_{e}}}
\newcommand{\jfbm}{\mathrm{j^{fb}_{\mu}}}
\newcommand{\jfbt}{\mathrm{j^{fb}_{\tau}}}
\newcommand{\jfbl}{\mathrm{j^{fb}_{\ell}}}

\newcommand{\cost}{\ensuremath{\cos\theta}}

\begin{document}
\begin{titlepage}
  \title{\boldmath
    Determination of $\gamma$/Z interference \\
    in {\EE} annihilation at LEP}
  \author{{\Large L3 Collaboration}}
  \begin{abstract}
    An S--Matrix ansatz is used to determine the mass and width of the Z boson, as
    well as the contributions of $\gamma$/Z interference and Z boson exchange to
    fermion--pair production.  For this purpose we use hadron and lepton--pair
    production cross sections and lepton forward--backward asymmetries that have
    been measured with the L3 detector at centre--of--mass energies between
    87~\GeV\ and 189~\GeV.
  \end{abstract}
\submitted
\end{titlepage}

\section*{Introduction}
The successful operation of LEP has allowed a precise measurement of
fermion--pair production, {\EEFF}, at centre--of--mass energies near the Z
resonance. The mass and the total and partial widths of the Z boson have been
determined with excellent accuracy. The Standard
Model\cite{standard_model,Veltman_SM} is confirmed with high precision by the
experimental results of L3\cite{l3-202} and other experiments\cite{aleph-216,
delphi-bible,
opal-86,
sld-05}. In these analyses, the contribution
of the interference between the photon and the Z boson exchange amplitudes is
fixed to the Standard Model expectation.

In this paper, we use an S--Matrix\cite{SMATRIX} approach to determine the
contributions of $\gamma$/Z interference and Z boson exchange, thus reducing the
number of theoretical assumptions. At centre--of--mass energies well above the Z
resonance, the reduced importance of Z boson exchange allows the determination, in
particular, of $\gamma$/Z interference with enhanced precision. The running of
LEP in 1997 and 1998 at energies of 182.7~{\GeV} and 188.7~{\GeV}, together with
a tenfold increase of integrated luminosity at high energy compared to our
previous S--Matrix analysis\cite{l3-117}, improves substantially the sensitivity
to the S--Matrix parameters. Similar analyses have been performed by the DELPHI
and OPAL collaborations\cite{delphi-186,opal-209}.

\section*{Measurements of Fermion--Pair Production}

Measurements of cross sections and forward--backward asymmetries for the
reactions {\EEFF}, have been performed with the L3
detector\cite{l3-00} at
centre--of--mass energies, $\sqrt{s}$, in the vicinity of the Z
resonance\cite{l3-202} and at 130.0, 136.1, 161.3, 172.3, 182.7 and
188.7~{\GeV}\cite{l3-90,l3-117,l3-196}.

At energies greater than 130~{\GeV} both leptons of the {\EE} final state are
required to be in the range $44^{\circ}<\theta<136^{\circ}$, where $\theta$ is
the angle between the incoming electron and the outgoing lepton.  Muon- and
tau--pair candidates are selected by the cuts $|\cost|<0.9$ and $|\cost|<0.92$,
respectively, for both final--state leptons. Hadron events are selected in the
full solid angle. In total, 27470 hadron events and 9417 lepton--pair events are
selected. These correspond to an integrated luminosity of 258.7~pb$^{-1}$.

For centre--of--mass energies in the vicinity of the Z resonance, the
sensitivity to photon~exchange and $\gamma$/Z interference is suppressed due to
the large Z~exchange cross section. Therefore, a minimum effective
centre--of--mass energy, $\sqrt{s'}$, or a maximum acollinearity angle in the
Bhabha channel, are required to select events without substantial energy loss
due to initial state radiation.  The remaining event samples at $\sqrt{s}\ge
130$~{\GeV} contain, in total, 7785 hadron and 7704 lepton--pair events. The
details of the analyses such as selection procedures, efficiencies, background
contributions, measured cross sections and forward--backward asymmetries,
together with the statistical and systematic uncertainties, are discussed in
References \citen{l3-117}, \citen{l3-90}, and \citen{l3-196}.

\SM\ expectations are calculated using the ZFITTER\cite{ZFITTER} and
TOPAZ0\cite{TOPAZ0} programs with the following
values\cite{l3-202,l3-178,PDG,JEGERLEHNER,cdf-mtop-1}
for the Z boson mass, {\MZ}, the top quark mass,  {\Mt}, the Higgs boson mass,
{\MH}, the electromagnetic coupling constant, $\alpha$, and the strong coupling
constant $\alpha_{s}$:
\begin{equation}
 \begin{array}{r@{$\,=\,$}lr@{$\,=\,$}ll}
              \MZ & 91\,189.8 \pm 3.1 {\MeV}, &           
  1/\alpha(\MZ^2) & 128.887 \pm 0.089  , \\               
              \Mt & 173.8 \pm 5.2 {\GeV}, &               
\alpha_{s}(\MZ^2) & 0.119 \pm 0.002 , \\                  
\multicolumn{2}{r}{95.3 \GeV \leq \MH \leq 1 {\TeV}.} &   
\multicolumn{2}{r}{ }

 \end{array}
\label{eq:smpara}
\end{equation}

The results of our analysis are not sensitive to the uncertainties on these
parameters. At energies above the Z resonance, the theoretical uncertainties on
the predicted cross sections are estimated to be 0.5\%\cite{err-zf} except for
the predictions for large--angle Bhabha scattering which have a larger uncertainty of
2\%\cite{BHABHA-THEORY}. Uncertainties on the forward--backward asymmetries are
smaller and negligible compared to the statistical uncertainties of the measurements.
The theoretical uncertainties are propagated into the systematic uncertainty on our
results.

\section*{\boldmath Determination of $\gamma$/Z interference}
\label{sec:smatrix}

The measurements of cross sections and forward--backward asymmetries are
analysed in the framework of the S--Matrix ansatz.  The programs
SMATASY\cite{SMATASY}, together with ZFITTER and TOPAZ0, are used for the
calculation of the theoretical predictions and QED radiative corrections to the
cross sections and forward--backward asymmetries.

The lowest--order total cross section, $\mathrm{\sigma^0_{tot}}$, and
forward--backward asymmetry, $A^0_{\mathrm{fb}}$,\\
for $\EEFF$\cite{SMATRIX} are:
\begin{eqnarray}
\sigma^0_{a}(s) & = &
  \frac{4}{3}\pi\alpha^2
                         \left[
                               \frac{\gf^a}{s} +
                               \frac{\jf^a (s-\MZbar^2) + \rf^a \, s}
                                     {(s-\MZbar^2)^2 + \MZbar^2 \GZbar^2}
                         \right] ,
              \qquad\mathrm{where}~a=\mathrm{tot,fb} , \nonumber\\
A^0_{\mathrm{fb}}(s) & = &
  \frac{3}{4} \frac{\sigma^0_{\mathrm{fb}}(s)}{\sigma^0_{\mathrm{tot}}(s)} ,
  \qquad\mathrm{with}~\sigma^0_{\mathrm{fb}}=\frac{4}{3}\left(\sigma_\mathrm{f}-\sigma_\mathrm{b}\right).
\end{eqnarray}
The S--Matrix ansatz defines the Z resonance using a Breit--Wigner denominator
with an $s$--independent width. In other approaches, a Breit--Wigner denominator
with an $s$--dependent width is used, which implies the following transformation
of the values of the Z boson mass and width\cite{SMATRIX}: $\MZ =
\MZbar+34.1 \MeV$ and $\GZ = \GZbar+0.9 \MeV$. In the following, the fit results
are quoted after applying these transformations. The S--Matrix parameters $\rf$,
$\jf$ and $\gf$ give the Z exchange, $\gamma$/Z interference and photon
exchange contributions for fermions of type $f$, respectively. For hadronic final
states the parameters $\rtoth$, $\jtoth$ and $\gtoth$ are sums over all produced quark
flavours.

\begin{figure}[htbp]
  \begin{center}
    \includegraphics[width=.9\textwidth]{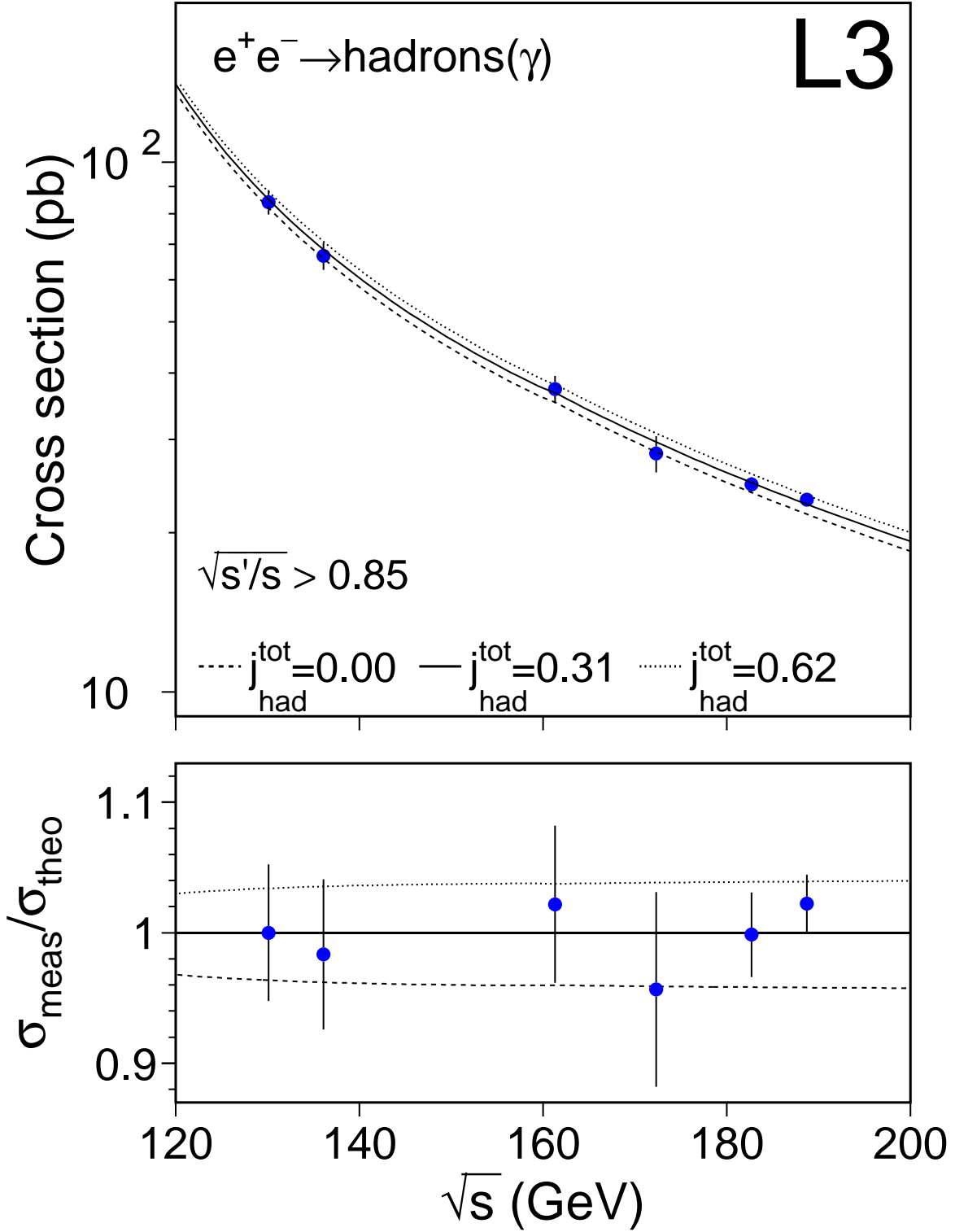}
    \parbox{\capwidth}{
      \caption[]{
        Cross sections for the process {$\epem\rightarrow\mbox{hadrons}(\gamma)$},
        for $\sqrt{s'/s} > 0.85$.
        The lines represent the theory predictions for different values of $\jtoth$.
        The lower plot normalises the measurements and predictions to that of
        $\jtoth = 0.31$.}
      \label{fig:ha_xsec}}
  \end{center}
\end{figure}

In Figure~\ref{fig:ha_xsec} the measured hadronic cross sections are compared to
the theory predictions for different values of $\jtoth$.  The S--Matrix
parameters are determined in a $\chi^2$ fit to our published
measurements\cite{l3-196} at centre--of--mass energies of 130 to 189~\GeV, using
our results of S--Matrix fits to the Z--peak data\cite{l3-202} as constraints.
As a cross--check, a fit to all cross--section and asymmetry measurements is
performed, which gives identical results. Correlations between measurements taken
close to the Z resonance and measurements at high centre--of--mass energies are
estimated and their influence on the fit results is found to be
negligible. Correlations between different measurements at high centre--of--mass
energy are taken into account in the fits.  The photon exchange
contributions, $\gf$, are fixed in the fits to their QED predictions. The fit results do not
depend on the uncertainty on $\alpha(\MZ^2)$.

The fitted S--Matrix parameters for electrons, muons, taus and hadrons, and
their correlations, are listed in Tables~\ref{tab:smat_pars},~\ref{tab:smat_corr8}
and~\ref{tab:smat_corr16}. The fits are performed with and without the assumption
of lepton universality. The parameters obtained for the individual lepton types are
compatible with each other and support this assumption.

\begin{figure}[htbp]
  \begin{center}
    \includegraphics[width=0.9\textwidth]{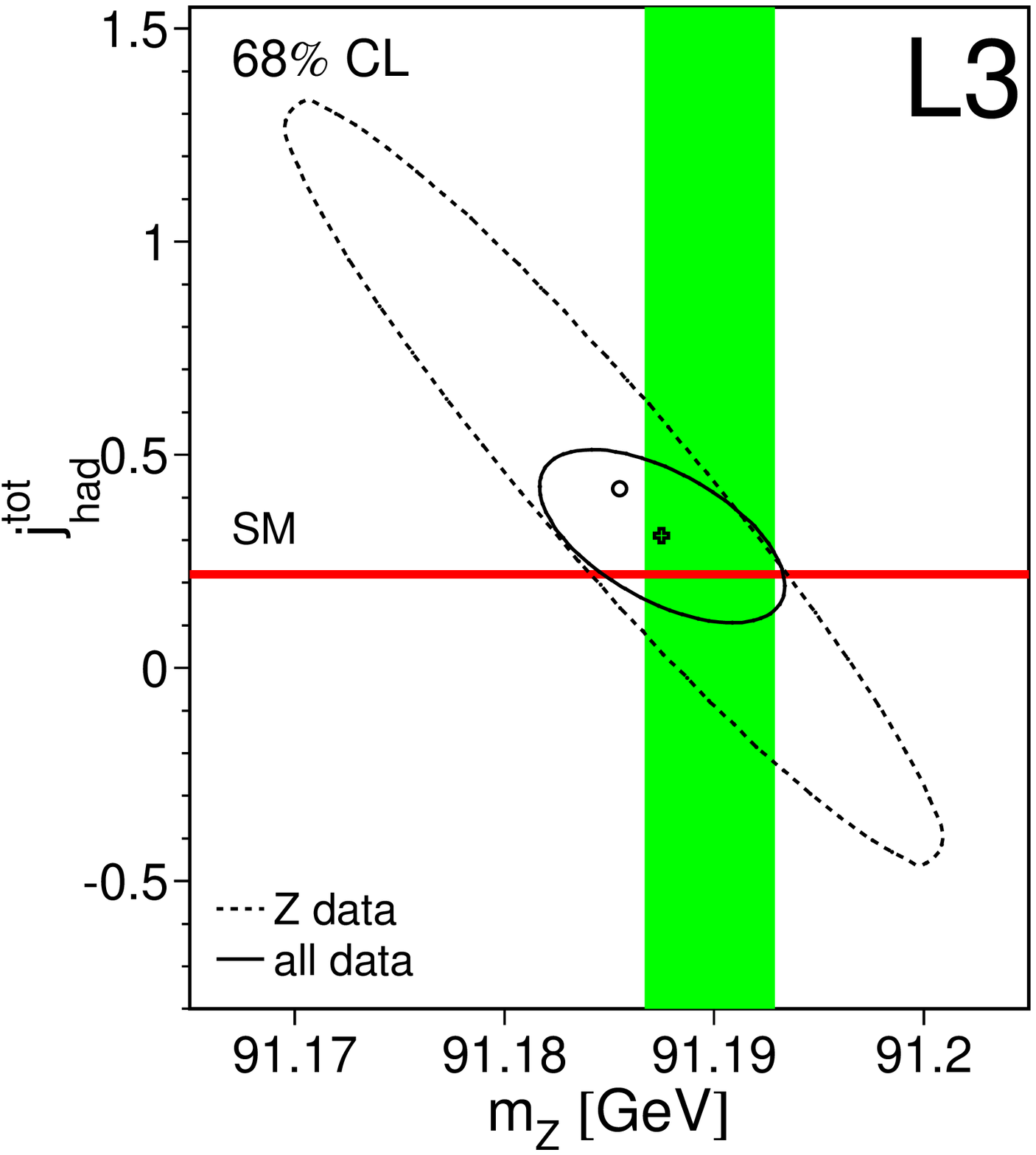}
    \parbox{\capwidth}{
      \caption[]{
        Contours in the ($\MZ$, $\jtoth$) plane at 68\% confidence level under
        the assumption of lepton universality. The dashed line is obtained from Z data
        only; the inclusion of 130~{\GeV} to 189~{\GeV} data gives the solid line.
        The circle (Z data) and the cross (all data) indicate the
        central values of the fits. The {\SM} prediction for $\jtoth$ is shown as the
        horizontal band. The vertical band corresponds to the 68\% confidence
        level interval on {\MZ} in a fit assuming the Standard Model value for $\gamma$/Z
        interference.}
      \label{fig:mz_jhad}}
  \end{center}
\end{figure}

A large correlation is observed between the mass of the Z boson and the
hadronic $\gamma$/Z interference term, $\jtoth$. This correlation causes an
increase in the uncertainty on $\MZ$ with respect to fits where the hadronic $\gamma$/Z
interference term is fixed to its {\SM} prediction. The value of this
correlation is reduced from -0.95 when using Z peak data alone\cite{l3-202} to -0.57.
Under the assumption of lepton universality the fitted hadronic $\gamma$/Z
interference term is:
\begin{eqnarray}
  \jtoth & = & 0.31 \pm 0.13 (\mathrm{exp}) \pm 0.04 (\mathrm{th}) \,.
  \nonumber
\end{eqnarray}
This value agrees with the {\SM} prediction of 0.21 and improves significantly the
precision of our previous S--Matrix analyses\cite{l3-202,l3-117}.  The theoretical
uncertainty is due to the uncertainty of 0.5\% on the calculation of cross
sections for $\sqrt{s} >$ 130~{\GeV} of the ZFITTER program. For the leptonic
$\gamma$/Z interference terms we obtain similar improvements.  The fitted value
for {\MZ} is:
\begin{eqnarray}
  \MZ & = & 91\,187.5 \pm 3.1(\mathrm{exp}) \pm 2.3(\Delta\jtoth) \pm
  0.6(\mathrm{th}) \MeV \,.
  \nonumber
\end{eqnarray}
The contribution to the uncertainty due to the $\gamma$/Z interference is separated
from the rest of the experimental uncertainty.  It is reduced significantly with
respect to our previous results\cite{l3-202,l3-117}. Again, the theoretical
uncertainty is due to the limited precision of the cross section calculations with
the ZFITTER program.  Figure~\ref{fig:mz_jhad} shows the 68\% confidence level
contours in the ($\MZ$, $\jtoth$) plane for the data taken at the Z--pole and
after including the 130--189~{\GeV} measurements. The improvement resulting from the
inclusion of the high energy measurements is clearly visible.

Our measurement of the hadronic interference term agrees with results from data
taken at energies below the Z resonance\cite{jhad_Topaz}. Adding
these low energy measurements to the fits, we obtain:
\begin{eqnarray*}
  \jtoth & = & 0.159 \pm 0.082 (\mathrm{exp}) \pm 0.015 (\mathrm{th}) \,,\\
  \MZ & = & 91\,190.4 \pm 3.1(\mathrm{exp}) \pm 1.5(\Delta\jtoth) \pm
  0.3(\mathrm{th}) \MeV \,,
\end{eqnarray*}
with a correlation coefficient of -0.40 between these quantities.

In summary, we use the S--Matrix framework to analyse data taken at the Z
resonance and at higher centre--of--mass energies. Our measurements
provide an improved determination of the $\gamma$/Z interference terms and the
Z boson mass. All parameters show good agreement with the Standard Model
predictions.

\section*{Acknowledgements}

We wish to express our gratitude to the CERN accelerator divisions for the
excellent performance of the LEP machine.  We acknowledge the contributions of
the engineers and technicians who have participated in the construction and
maintenance of this experiment.

\clearpage
\bibliographystyle{l3stylem}

\clearpage
\typeout{   }     
\typeout{Using author list for paper 214 -- ? }
\typeout{$Modified: Tue Jun 27 13:18:16 2000 by clare $}
\typeout{!!!!  This should only be used with document option a4p!!!!}
\typeout{   }
%
%
%
%
%
%

\newcount\tutecount  \tutecount=0
\def\tutenum#1{\global\advance\tutecount by 1 \xdef#1{\the\tutecount}}
\def\tute#1{$^{#1}$}
\tutenum\aachen            
\tutenum\nikhef            
\tutenum\mich              
\tutenum\lapp              
\tutenum\basel             
\tutenum\lsu               
\tutenum\beijing           
\tutenum\berlin            
\tutenum\bologna           
\tutenum\tata              
\tutenum\ne                
\tutenum\bucharest         
\tutenum\budapest          
\tutenum\mit               
\tutenum\debrecen          
\tutenum\florence          
\tutenum\cern              
\tutenum\wl                
\tutenum\geneva            
\tutenum\hefei             
\tutenum\seft              
\tutenum\lausanne          
\tutenum\lecce             
\tutenum\lyon              
\tutenum\madrid            
\tutenum\milan             
\tutenum\moscow            
\tutenum\naples            
\tutenum\cyprus            
\tutenum\nymegen           
\tutenum\caltech           
\tutenum\perugia           
\tutenum\cmu               
\tutenum\prince            
\tutenum\rome              
\tutenum\peters            
\tutenum\potenza           
\tutenum\salerno           
\tutenum\ucsd              
\tutenum\santiago          
\tutenum\sofia             
\tutenum\korea             
\tutenum\alabama           
\tutenum\utrecht           
\tutenum\purdue            
\tutenum\psinst            
\tutenum\zeuthen           
\tutenum\eth               
\tutenum\hamburg           
\tutenum\taiwan            
\tutenum\tsinghua          

{
\parskip=0pt
\noindent
{\bf The L3 Collaboration:}
\ifx\selectfont\undefined
 \baselineskip=10.8pt
 \baselineskip\baselinestretch\baselineskip
 \normalbaselineskip\baselineskip
 \ixpt
\else
 \fontsize{9}{10.8pt}\selectfont
\fi
\medskip
\tolerance=10000
\hbadness=5000
\raggedright
\hsize=162truemm\hoffset=0mm
\def\r{\rlap,}
\noindent

M.Acciarri\r\tute\milan\
P.Achard\r\tute\geneva\ 
O.Adriani\r\tute{\florence}\ 
M.Aguilar-Benitez\r\tute\madrid\ 
J.Alcaraz\r\tute\madrid\ 
G.Alemanni\r\tute\lausanne\
J.Allaby\r\tute\cern\
A.Aloisio\r\tute\naples\ 
M.G.Alviggi\r\tute\naples\
G.Ambrosi\r\tute\geneva\
% ...
\newpage
\begin{list}{A}{\itemsep=0pt plus 0pt minus 0pt\parsep=0pt plus 0pt minus 0pt
                \topsep=0pt plus 0pt minus 0pt}
\item[\aachen]
 I. Physikalisches Institut, RWTH, D-52056 Aachen, FRG$^{\S}$\\
 III. Physikalisches Institut, RWTH, D-52056 Aachen, FRG$^{\S}$
\item[\nikhef] National Institute for High Energy Physics, NIKHEF, 
     and University of Amsterdam, NL-1009 DB Amsterdam, The Netherlands
\item[\mich] University of Michigan, Ann Arbor, MI 48109, USA
\item[\lapp] Laboratoire d'Annecy-le-Vieux de Physique des Particules, 
     LAPP,IN2P3-CNRS, BP 110, F-74941 Annecy-le-Vieux CEDEX, France
\item[\basel] Institute of Physics, University of Basel, CH-4056 Basel,
     Switzerland
\item[\lsu] Louisiana State University, Baton Rouge, LA 70803, USA
\item[\beijing] Institute of High Energy Physics, IHEP, 
  100039 Beijing, China$^{\triangle}$ 
\item[\berlin] Humboldt University, D-10099 Berlin, FRG$^{\S}$
\item[\bologna] University of Bologna and INFN-Sezione di Bologna, 
     I-40126 Bologna, Italy
\item[\tata] Tata Institute of Fundamental Research, Bombay 400 005, India
\item[\ne] Northeastern University, Boston, MA 02115, USA
\item[\bucharest] Institute of Atomic Physics and University of Bucharest,
     R-76900 Bucharest, Romania
\item[\budapest] Central Research Institute for Physics of the 
     Hungarian Academy of Sciences, H-1525 Budapest 114, Hungary$^{\ddag}$
\item[\mit] Massachusetts Institute of Technology, Cambridge, MA 02139, USA
\item[\debrecen] KLTE-ATOMKI, H-4010 Debrecen, Hungary$^\P$
\item[\florence] INFN Sezione di Firenze and University of Florence, 
     I-50125 Florence, Italy
\item[\cern] European Laboratory for Particle Physics, CERN, 
     CH-1211 Geneva 23, Switzerland
\item[\wl] World Laboratory, FBLJA  Project, CH-1211 Geneva 23, Switzerland
\item[\geneva] University of Geneva, CH-1211 Geneva 4, Switzerland
\item[\hefei] Chinese University of Science and Technology, USTC,
      Hefei, Anhui 230 029, China$^{\triangle}$
\item[\seft] SEFT, Research Institute for High Energy Physics, P.O. Box 9,
      SF-00014 Helsinki, Finland
\item[\lausanne] University of Lausanne, CH-1015 Lausanne, Switzerland
\item[\lecce] INFN-Sezione di Lecce and Universit\'a Degli Studi di Lecce,
     I-73100 Lecce, Italy
\item[\lyon] Institut de Physique Nucl\'eaire de Lyon, 
     IN2P3-CNRS,Universit\'e Claude Bernard, 
     F-69622 Villeurbanne, France
\item[\madrid] Centro de Investigaciones Energ{\'e}ticas, 
     Medioambientales y Tecnolog{\'\i}cas, CIEMAT, E-28040 Madrid,
     Spain${\flat}$ 
\item[\milan] INFN-Sezione di Milano, I-20133 Milan, Italy
\item[\moscow] Institute of Theoretical and Experimental Physics, ITEP, 
     Moscow, Russia
\item[\naples] INFN-Sezione di Napoli and University of Naples, 
     I-80125 Naples, Italy
\item[\cyprus] Department of Natural Sciences, University of Cyprus,
     Nicosia, Cyprus
\item[\nymegen] University of Nijmegen and NIKHEF, 
     NL-6525 ED Nijmegen, The Netherlands
\item[\caltech] California Institute of Technology, Pasadena, CA 91125, USA
\item[\perugia] INFN-Sezione di Perugia and Universit\'a Degli 
     Studi di Perugia, I-06100 Perugia, Italy   
\item[\cmu] Carnegie Mellon University, Pittsburgh, PA 15213, USA
\item[\prince] Princeton University, Princeton, NJ 08544, USA
\item[\rome] INFN-Sezione di Roma and University of Rome, ``La Sapienza",
     I-00185 Rome, Italy
\item[\peters] Nuclear Physics Institute, St. Petersburg, Russia
\item[\potenza] INFN-Sezione di Napoli and University of Potenza, 
     I-85100 Potenza, Italy
\item[\salerno] University and INFN, Salerno, I-84100 Salerno, Italy
\item[\ucsd] University of California, San Diego, CA 92093, USA
\item[\santiago] Dept. de Fisica de Particulas Elementales, Univ. de Santiago,
     E-15706 Santiago de Compostela, Spain
\item[\sofia] Bulgarian Academy of Sciences, Central Lab.~of 
     Mechatronics and Instrumentation, BU-1113 Sofia, Bulgaria
\item[\korea]  Laboratory of High Energy Physics, 
     Kyungpook National University, 702-701 Taegu, Republic of Korea
\item[\alabama] University of Alabama, Tuscaloosa, AL 35486, USA
\item[\utrecht] Utrecht University and NIKHEF, NL-3584 CB Utrecht, 
     The Netherlands
\item[\purdue] Purdue University, West Lafayette, IN 47907, USA
\item[\psinst] Paul Scherrer Institut, PSI, CH-5232 Villigen, Switzerland
\item[\zeuthen] DESY, D-15738 Zeuthen, 
     FRG
\item[\eth] Eidgen\"ossische Technische Hochschule, ETH Z\"urich,
     CH-8093 Z\"urich, Switzerland
\item[\hamburg] University of Hamburg, D-22761 Hamburg, FRG
\item[\taiwan] National Central University, Chung-Li, Taiwan, China
\item[\tsinghua] Department of Physics, National Tsing Hua University,
      Taiwan, China
\item[\S]  Supported by the German Bundesministerium 
        f\"ur Bildung, Wissenschaft, Forschung und Technologie
\item[\ddag] Supported by the Hungarian OTKA fund under contract
numbers T019181, F023259 and T024011.
\item[\P] Also supported by the Hungarian OTKA fund under contract
  numbers T22238 and T026178.
\item[$\flat$] Supported also by the Comisi\'on Interministerial de Ciencia y 
        Tecnolog{\'\i}a.
\item[$\sharp$] Also supported by CONICET and Universidad Nacional de La Plata,
        CC 67, 1900 La Plata, Argentina.
\item[$\diamondsuit$] Also supported by Panjab University, Chandigarh-160014, 
        India.
\item[$\triangle$] Supported by the National Natural Science
  Foundation of China.
\end{list}
}
\vfill


\clearpage

\begin{table}[ht] 
  \begin{center}
    \renewcommand{\arraystretch}{1.1}
    \begin{tabular}{| c | r@{$ \, \pm \, $}l | r@{$ \, \pm \, $}l | c | c |}
\hline
Parameter      &            \mfcl{Treatment of Charged Leptons} & Theory      & Standard \\
               & \mtcl{Non--Universality} & \mtcl{Universality} & uncertainty &   Model  \\
\hline
\hline
$\MZ$~~[\MeV]  &         91188.3 &    3.9 &   91187.5 &     3.9 & 0.6         &      --- \\
$\GZ$~~[\MeV]  &          2502.8 &    4.1 &    2502.5 &     4.1 & 0.1         &   $2492.7\,^{+3.8}_{-5.2}$ \\
\hline                                                        
$\rtoth$       &          2.9856 & 0.0092 &    2.9848 &  0.0092 & 0.0003      &   $2.9584\,^{+0.0088}_{-0.0119}$ \\
$\rtote$       &         0.14317 & 0.00075&          \mtcl{---} & 0.00002     &          \\
$\rtotm$       &         0.14287 & 0.00079&          \mtcl{---} & 0.00002     &          \\
$\rtott$       &         0.14375 & 0.00102&          \mtcl{---} & 0.00002     &          \\
$\rtotl$       &               \mtcl{---} &   0.14318 & 0.00059 & 0.00002     &  $0.14242\,^{+0.00035}_{-0.00049}$ \\
\hline                                                        
$\jtoth$       &            0.30 &   0.13 &      0.31 &    0.13 & 0.04        &     $0.21 \pm 0.01 $ \\
$\jtote$       &         --0.030 &  0.045 &          \mtcl{---} & 0.002       &          \\
$\jtotm$       &         --0.001 &  0.027 &          \mtcl{---} & 0.002       &          \\
$\jtott$       &           0.061 &  0.031 &          \mtcl{---} & 0.002       &          \\
$\jtotl$       &               \mtcl{---} &     0.017 &   0.019 & 0.002       &    $0.0041 \pm 0.0003$ \\
\hline                                     
$\rfbe$        &         0.00177 & 0.00111&          \mtcl{---} & 0.000002    &          \\
$\rfbm$        &         0.00333 & 0.00064&          \mtcl{---} & 0.000002    &          \\
$\rfbt$        &         0.00448 & 0.00092&          \mtcl{---} & 0.000002    &          \\
$\rfbl$        &               \mtcl{---} &   0.00332 & 0.00047 & 0.000002    &  $0.00255 \pm 0.00023$ \\
\hline                                     
$\jfbe$        &           0.700 &  0.075 &          \mtcl{---} & 0.001       &          \\
$\jfbm$        &           0.807 &  0.034 &          \mtcl{---} & 0.001       &          \\
$\jfbt$        &           0.732 &  0.044 &          \mtcl{---} & 0.001       &          \\
$\jfbl$        &               \mtcl{---} &     0.770 &   0.026 & 0.001       &   $0.799 \pm 0.001 $ \\
\hline
$\chi^2$ /d.o.f. &           \mtcl{30.4/28} &      \mtcl{33.0/36} &&      --- \\
\hline
    \end{tabular} \vskip 0.5cm
    \parbox{\capwidth}{
      \caption[]{
        Results of the fits in the S--Matrix framework without and with the
        assumption of lepton universality. The theory uncertainties on the
        S--Matrix parameters are determined from the 0.5\% uncertainty on the
        ZFITTER predictions for cross sections.  The {\SM} expectations are
        calculated using the parameters listed in Equation~\ref{eq:smpara}.}
      \label{tab:smat_pars}}
  \end{center}
\end{table}

\begin{table}
 \begin{sideways}
 \begin{minipage}[b]{\textheight}
 \ \vspace*{10mm} \\
 \small
 \begin{center}
 \renewcommand{\arraystretch}{1.1}
 \begin{tabular}{|l|rrrrrrrrrrrrrrrr|}
 \hline
        &  $\MZ$  &   $ \GZ$ & $\rtoth $
                               & $\rtote$ 
                                       & $\rtotm$ 
                                               & $\rtott$  
                                                       & $\jtoth$ 
                                                               & $\jtote$ 
                                                                       & $\jtotm$        
                                                                               & $\jtott$ 
                                                                                       & $\rfbe$        
                                                                                               & $\rfbm$        
                                                                                                       & $\rfbt$        
                                                                                                               & $\jfbe$
                                                                                                                       & $\jfbm$        
                                                                                                                               &$\jfbt$ \\
 \hline
$\MZ$   &$\phantom{-}1.00$
                &$\phantom{-}0.38$
                        &$\phantom{-}0.07$
                                &$-0.08$&$\phantom{-}0.01$
                                                &$\phantom{-}0.01$
                                                        &$-0.58$&$-0.14$&$-0.15$&$-0.14$&$-0.05$&$\phantom{-}0.08$
                                                                                                        &$\phantom{-}0.03$
                                                                                                                &$-0.01$&$-0.06$&$-0.02$\\
$\GZ$   &$     $&$ 1.00$&$ 0.91$&$ 0.54$&$ 0.52$&$ 0.40$&$ 0.01$&$-0.02$&$ 0.02$&$ 0.02$&$ 0.00$&$ 0.02$&$ 0.02$&$-0.01$&$ 0.04$&$ 0.04$\\
$\rtoth$&$     $&$     $&$ 1.00$&$ 0.56$&$ 0.53$&$ 0.41$&$ 0.01$&$-0.03$&$ 0.01$&$ 0.01$&$ 0.00$&$ 0.02$&$ 0.02$&$-0.01$&$ 0.04$&$ 0.04$\\
$\rtote$&$     $&$     $&$     $&$ 1.00$&$ 0.33$&$ 0.25$&$ 0.08$&$ 0.05$&$ 0.04$&$ 0.03$&$ 0.13$&$-0.01$&$ 0.00$&$-0.02$&$ 0.04$&$ 0.02$\\
$\rtotm$&$     $&$     $&$     $&$     $&$ 1.00$&$ 0.23$&$ 0.00$&$-0.02$&$ 0.10$&$ 0.02$&$ 0.00$&$ 0.03$&$ 0.01$&$-0.01$&$ 0.11$&$ 0.02$\\
$\rtott$&$     $&$     $&$     $&$     $&$     $&$ 1.00$&$ 0.00$&$-0.01$&$ 0.02$&$ 0.09$&$ 0.00$&$ 0.01$&$ 0.03$&$-0.01$&$ 0.02$&$ 0.10$\\
$\jtoth$&$     $&$     $&$     $&$     $&$     $&$     $&$ 1.00$&$ 0.13$&$ 0.13$&$ 0.13$&$ 0.04$&$-0.04$&$-0.03$&$ 0.01$&$ 0.05$&$ 0.02$\\
$\jtote$&$     $&$     $&$     $&$     $&$     $&$     $&$     $&$ 1.00$&$ 0.03$&$ 0.03$&$ 0.00$&$-0.01$&$-0.01$&$ 0.30$&$ 0.01$&$ 0.01$\\
$\jtotm$&$     $&$     $&$     $&$     $&$     $&$     $&$     $&$     $&$ 1.00$&$ 0.03$&$ 0.01$&$ 0.07$&$-0.01$&$ 0.00$&$ 0.31$&$ 0.01$\\
$\jtott$&$     $&$     $&$     $&$     $&$     $&$     $&$     $&$     $&$     $&$ 1.00$&$ 0.01$&$-0.01$&$ 0.04$&$ 0.00$&$ 0.01$&$ 0.15$\\
$\rfbe$ &$     $&$     $&$     $&$     $&$     $&$     $&$     $&$     $&$     $&$     $&$ 1.00$&$-0.01$&$ 0.00$&$ 0.03$&$ 0.00$&$ 0.00$\\
$\rfbm$ &$     $&$     $&$     $&$     $&$     $&$     $&$     $&$     $&$     $&$     $&$     $&$ 1.00$&$ 0.01$&$ 0.00$&$ 0.13$&$ 0.00$\\
$\rfbt$ &$     $&$     $&$     $&$     $&$     $&$     $&$     $&$     $&$     $&$     $&$     $&$     $&$ 1.00$&$ 0.00$&$ 0.00$&$ 0.11$\\
$\jfbe$ &$     $&$     $&$     $&$     $&$     $&$     $&$     $&$     $&$     $&$     $&$     $&$     $&$     $&$ 1.00$&$ 0.00$&$ 0.00$\\
$\jfbm$ &$     $&$     $&$     $&$     $&$     $&$     $&$     $&$     $&$     $&$     $&$     $&$     $&$     $&$     $&$ 1.00$&$ 0.00$\\
$\jfbt$ &$     $&$     $&$     $&$     $&$     $&$     $&$     $&$     $&$     $&$     $&$     $&$     $&$     $&$     $&$     $&$ 1.00$\\
 \hline
 \end{tabular}
 \renewcommand{\arraystretch}{1.0}
 \parbox{\capwidth}{
 \caption[]{Correlation coefficients of the S--Matrix parameters listed in 
            Table~\protect\ref{tab:smat_pars} 
            not assuming lepton universality.
            } 
   \label{tab:smat_corr16}
   }
 \end{center}
 \normalsize
 \end{minipage}
 \end{sideways}
\end{table}

\begin{table}[hb]
  \renewcommand{\arraystretch}{1.1}
  \begin{center}
    \begin{tabular}{|l|r|r|r|r|r|r|r|r|}
\hline
&    \mocl{$\MZ$} & \mocl{$\GZ$}
& \mocl{$\rtoth$} & \mocl{$\rtotl$}
& \mocl{$\jtoth$} & \mocl{$\jtotl$}
&  \mocl{$\rfbl$} &  \mocl{$\jfbl$} \\
\hline
\hline
$\MZ$    &  1.00 &  0.05 &  0.06 &--0.02 &--0.57 &--0.24 &  0.05 &--0.06 \\
$\GZ$    &       &  1.00 &  0.92 &  0.69 &  0.01 &  0.01 &  0.02 &  0.05 \\
$\rtoth$ &       &       &  1.00 &  0.71 &  0.01 &  0.00 &  0.03 &  0.05 \\
$\rtotl$ &       &       &       &  1.00 &  0.04 &  0.08 &  0.05 &  0.08 \\
$\jtoth$ &       &       &       &       &  1.00 &  0.21 &--0.03 &  0.06 \\
$\jtotl$ &       &       &       &       &       &  1.00 &  0.04 &  0.25 \\
$\rfbl $ &       &       &       &       &       &       &  1.00 &  0.11 \\
$\jfbl $ &       &       &       &       &       &       &       &  1.00 \\
\hline
    \end{tabular} \vskip 0.5cm
    \parbox{\capwidth}{
      \caption[]{Correlation coefficients of the S--Matrix parameters listed in 
        Table~\protect\ref{tab:smat_pars} 
        assuming lepton universality.
        } 
      \label{tab:smat_corr8}}
  \end{center}
\end{table}

\clearpage

\end{document}